\documentclass[fleqn,usenatbib]{rasti}

\usepackage[T1]{fontenc}

\DeclareRobustCommand{\VAN}[3]{#2}
\let\VANthebibliography\thebibliography
\def\thebibliography{\DeclareRobustCommand{\VAN}[3]{##3}\VANthebibliography}

\usepackage{graphicx}	
\usepackage{amsmath}	
\usepackage{amssymb}    

\usepackage{bm}
\usepackage{xcolor}
\usepackage{algorithm}
\usepackage{algorithmic}
\usepackage{siunitx}
\usepackage{float}
\usepackage[switch]{lineno}
\usepackage{etoolbox} 

\newcommand*\linenomathpatch[1]{%
  \cspreto{#1}{\linenomath}%
  \cspreto{#1*}{\linenomath}%
  \csappto{end#1}{\endlinenomath}%
  \csappto{end#1*}{\endlinenomath}%
}
\newcommand*\linenomathpatchAMS[1]{%
  \cspreto{#1}{\linenomathAMS}%
  \cspreto{#1*}{\linenomathAMS}%
  \csappto{end#1}{\endlinenomath}%
  \csappto{end#1*}{\endlinenomath}%
}

\expandafter\ifx\linenomath\linenomathWithnumbers
  \let\linenomathAMS\linenomathWithnumbers
  \patchcmd\linenomathAMS{\advance\postdisplaypenalty\linenopenalty}{}{}{}
\else
  \let\linenomathAMS\linenomathNonumbers
\fi

\linenomathpatch{equation}
\linenomathpatchAMS{gather}
\linenomathpatchAMS{multline}
\linenomathpatchAMS{align}
\linenomathpatchAMS{alignat}
\linenomathpatchAMS{flalign}

\usepackage{newtxtext,newtxmath}

\newcommand{\STWO}{\ensuremath{\mathbb{S}^2}}
\newcommand{\sothree}{\ensuremath{\mathrm{SO}(3)}}
\newcommand{\thetaphi}{\ensuremath{(\theta,\phi)}}
\newcommand{\thetaprimephiprime}{\ensuremath{(\theta',\phi')}}
\newcommand{\posterior}[1][]{\ensuremath{p(\bm{m}{#1}\bm{|d})}}
\newcommand{\likelihood}[1][]{\ensuremath{p(\bm{d|m}{#1})}}
\newcommand{\prior}[1][]{\ensuremath{p(\bm{m}{#1})}}

\DeclareMathOperator*{\argmin}{argmin}
\DeclareMathOperator*{\argmax}{argmax}

\title[Posterior sampling on the sphere in seismology and cosmology]{Posterior sampling for inverse imaging problems on the sphere in seismology and cosmology}

\author[A. Marignier et al.]{
Augustin Marignier,$^{1,2}$\thanks{E-mail: augustin.marignier.14@ucl.ac.uk}
Jason D. McEwen,$^{1}$
Ana M. G. Ferreira,$^{2,3}$ 
Thomas D. Kitching$^{1}$
\\
$^{1}$Mullard Space Science Laboratory, University College London, Dorking RH5 6NT, UK \\
$^{2}$Department of Earth Sciences, University College London, London WC1E 6BT, UK \\
$^{3}$CERIS, Instituto Superior T\'{e}cnico, Universidade de Lisboa, Lisbon, Portugal
}
\date{Accepted XXX. Received YYY; in original form ZZZ}

\pubyear{2022}

\begin{document}\label{firstpage}
\pagerange{\pageref{firstpage}--\pageref{lastpage}}
\maketitle

\begin{abstract}
    In this work, we describe a framework for solving spherical inverse imaging problems using posterior sampling for full uncertainty quantification.
    Inverse imaging problems defined on the sphere arise in many fields, including seismology and cosmology where images are defined on the globe and the cosmic sphere, and are generally high-dimensional and computationally expensive.
    As a result, sampling the posterior distribution of spherical imaging problems is a challenging task.
    Our framework leverages a proximal Markov chain Monte Carlo (MCMC) algorithm to efficiently sample the high-dimensional space of spherical images with a sparsity-promoting wavelet prior.
    We detail the modifications needed for the algorithm to be applied to spherical problems, and give special consideration to the crucial forward modelling step which contains computationally expensive spherical harmonic transforms.
    By sampling the posterior, our framework allows for full and flexible uncertainty quantification, something which is not possible with other methods based on, for example, convex optimisation.
    We demonstrate our framework in practice on full-sky cosmological mass-mapping and to the construction of phase velocity maps in global seismic tomography.
    We find that our approach is potentially useful at moderate resolutions, such as those of interest in seismology.
    However at high resolutions, such as those required for astrophysical applications, the poor scaling of the complexity of spherical harmonic transforms severely limits our method, which may be resolved with future GPU implementations.
    A new Python package, \texttt{pxmcmc}, containing the proximal MCMC sampler, measurement operators, wavelet transforms and sparse priors is made publicly available.
\end{abstract}




\section{Introduction}
Inverse problems on the sphere are common in many fields, from astrophysics \citep[e.g.][]{Planck2015,DESY3}, to geophysics \citep[e.g.][]{Ritsema2011,Chang2015} and more.
These tend to be high-dimensional and computationally challenging imaging problems, increasingly so at high resolutions.
Spherical inverse problems are difficult to solve using posterior sampling methods such as Markov chain Monte Carlo (MCMC), largely due to the generally large number of parameters to sample and the high computational cost of repeated evaluations of the forward problem on the sphere.
MCMC methods have grown in popularity in recent decades \citep[e.g.][]{Mosegaard1995,Maliverno2002,Bodin2012,Lewis2002,Corless2009,Cai2018I}, benefitting from their ability to sample the full posterior probability density function (pdf), which constitutes the solution of the inverse problem.
This allows for flexible calculation of any measure of uncertainty in the solution.
Furthermore, they can be used to solve non-linear inverse problems, which are commonplace in, for example, geophysics \citep{Maliverno2002, Bodin2012, Ferreira2020}.
However, sampling methods come at significant computational cost, for both solving the inverse problem and for model comparison, which is often computationally infeasible.
The simpler MCMC methods, such as the Metropolis-Hastings (MH) algorithm, are known to struggle in high-dimensional parameter spaces \citep[e.g.][]{Roberts1998,Neal2012}, such as those for spherical imaging problems.
The exponential increase in volume of the parameter space makes MH unable to converge to a solution.
Alternative gradient-based methods such as the Hamiltonian Monte Carlo (HMC) \citep{Duane1987,Neal2012} and the unadjusted Langevin algorithm (ULA) \citep{Roberts1996} scale favourably compared to the MH algorithm as the number of dimensions increases \citep{Roberts1998,Neal2012}.
The caveat here is that these methods can only be applied to smooth distributions, limiting the form of prior information that can be used as regularisation.
Conversely, nested sampling methods \citep{Skilling2006} can be used for non-smooth problems.
However, most nested sampling approaches cannot scale to the high-dimensional settings of inverse problems without gradient information.
Proximal MCMC methods have recently been developed \citep{Pereyra2016}, leveraging proximal mappings to scale to high dimensions and allow for non-smooth distributions.
Proximal mappings have also recently been used to scale nested sampling methods to high-dimensional imaging problems \citep{Cai2022}.

Compressed sensing \citep{Donoho2006,Candes2011} has demonstrated that sparse signals can be accurately recovered from incomplete data, recovering both sharp and smooth image features simultaneously from underdetermined systems \citep{Loris2007,Wallis2017}.
As a result, sparse priors have been widely adopted for solving inverse imaging problems, e.g.\ for radio interferometry applications \citep{Wiaux2009,Carrillo2014,Pratley2017,Cai2018I,Cai2018II}, cosmological mass-mapping \citep{Lanusse2016,Price2020}, and have received some attention in seismic tomography \citep{Loris2007,Charlety2013}.
The prior pdf (i.e.\ the distribution representing \textit{a prior} beliefs to constrain the inverse problem) often used to promote sparsity is the Laplace distribution, which is non-differentiable and so cannot be used with gradient-based sampling algorithms.

Current approaches to uncertainty quantification for high-dimensional inverse problems with sparse a priori constraints exploit results from information theory to derive approximate credible regions from a single point estimate solution of the inverse problem \citep{Pereyra2017,Cai2018II,Price2020}.
The point estimate solution can be obtained using convex optimisation \citep{Cai2018II,Price2020}, and thus this approach is much faster than posterior sampling, particularly for high-resolution spherical problems.
However, this information theory based approach can only approximate credible regions.
A sampling method on the other hand will have the flexibility to calculate any measure of uncertainty, as the samples represent a full probability distribution.

In this work we provide a framework for solving inverse imaging problems on the sphere using posterior sampling with sparsity-promoting priors.
For this, we leverage a proximal MCMC algorithm \citep{Pereyra2016}, which is a sampling scheme that uses proximal mappings to efficiently sample high-dimensional parameter spaces.
Proximal mappings can be viewed as more general gradient operators that can be used even on non-differentiable functions, allowing us to use a Laplace prior to promote sparsity.
In our framework we outline the modifications needed to the proximal MCMC algorithm for spherical problems, with particular consideration given to the spherical parameterisation and forward operator.
We then demonstrate our framework in practice first on a common problem from the field of global seismic tomography using both synthetic and real data, and then on a low-resolution full-sky cosmological mass-mapping example from simulation data.

This paper is structured as follows.
Section~\ref{sec:background} gives the necessary mathematical background for Bayesian inference and representations of spherical images.
Section~\ref{sec:framework} outlines our framework for posterior sampling of inverse problems on the sphere, with details of the proximal MCMC algorithm that we use.
Sections~\ref{sec:GST} and~\ref{sec:CMM} contain our illustrative examples from global seismic tomography and cosmological mass-mapping, respectively, and we conclude in Section~\ref{sec:concl}.

\section{Mathematical Background}\label{sec:background}
In this section we provide the necessary mathematical background for this work, including Bayesian inference for imaging, and harmonic and wavelet representations of spherical images.

\subsection{Bayesian inference for imaging}
Consider some observed data $\bm{d}$ and model parameters $\bm{m}$ that are related by some general, possibly non-linear, forward operator $\bm{G}$, as
\begin{equation}
    \bm{d}=\bm{G}(\bm{m}) + \bm{n},
    \label{eqn:general}
\end{equation}
where $\bm{n}$ represents observational noise.
The aim of the inverse problem is to infer $\bm{m}$ from $\bm{d}$.
This can be formulated in a Bayesian statistical framework by Bayes Theorem as 
\begin{equation}
    \posterior \propto \likelihood \prior.
    \label{eqn:bayes}
\end{equation}
Here \posterior\ is known as the posterior pdf that encapsulates knowledge of the model parameters $\bm{m}$ given data $\bm{d}$ and represents the solution to the inverse problem \citep{Tarantola1982,Mosegaard1995}.
The likelihood function $\likelihood$, assuming independent and identically distributed Gaussian noise with known standard deviation $\sigma$ in the data, is given by
\begin{equation}
    \likelihood \propto \mathrm{exp}\left(-\frac{\|\bm{d}-\bm{G}(\bm{m})\|_2^2}{2\sigma^2}\right).
\end{equation}
Inverse problems are typically ill-posed and often require some form of regularisation in the form of prior knowledge about the model parameters to further constrain the inverse problem.
In the Bayesian framework, this is represented by the prior pdf \prior, which can take many forms.
Common examples include the Uniform prior \citep{Jaynes2003}, $\prior \propto \mathrm{constant}$; the Gaussian prior \citep{Tarantola1982,Golub1999}, $\prior \propto \mathrm{exp}(-\mu\|\bm{m}\|_2^2)$; and the Laplace prior \citep{Donoho2006,Candes2011}, $\prior \propto \mathrm{exp}(-\mu\|\bm{m}\|_1)$, to promote sparsity.
In these priors, $\|\cdot\|_p$ is the $l_p$-norm and $\mu$ is a parameter that captures the width of the distribution.
Henceforth we use the Laplace prior as we wish to promote sparsity.
The constant of proportionality in equation~\ref{eqn:bayes} is known as the Bayesian evidence.
This is useful for comparing the physical model ($\bm{G}$) to alternative hypotheses, but may be ignored for inference about model parameters $\bm{m}$.

Solving the inverse problem by probabilistic sampling involves sampling the posterior pdf \posterior, most commonly using MCMC algorithms.
The simplest is the MH algorithm.
MH proposes a new sample from a proposal distribution, which has a certain probability of being accepted.
More sophisticated and better suited to high-dimensional problems than MH are gradient-based methods, such as the HMC and Langevin algorithms mentioned previously, which use the gradients of the target pdf to more efficiently guide the exploration of the parameter space.
While the set of samples from the posterior pdf represents the full solution of the inverse problem, results are generally reported in terms of summary statistics, such as the mean of the samples.
The \textit{maximum a posteriori} (MAP) solution,
\begin{equation}
    \bm{m}^{\mathrm{(MAP)}} = \argmax_{\bm{m}}\;\posterior
    \label{eqn:MAP}
\end{equation}
is also a common choice, however, due to the randomness of MCMC there is no guarantee of the MAP being found.
The highest posterior probability sample of the MCMC provides the closest estimate of the MAP, but can still be relatively far from the true MAP, particularly in higher dimensional spaces.
Additionally, some measure of uncertainty on individual model parameters can be calculated from the posterior samples.
We give further details of this in the next section.

In inverse imaging problems, the model parameters $\bm{m}$ represent a 2D (or possibly 3D) image.
MCMC can be used to sample either the direct pixel values $\bm{x}$ of the image, or the coefficients $\bm{\alpha}$ representing the image in a particular basis.
The forward operator $\bm{G}$ in equation~\ref{eqn:general} can be described as a (possibly non-linear) measurement operator $\bm{\Phi}$ acting on the image $\bm{x}$.
In many physical situations, the image $\bm{x}$ has a sparse representation in some basis, which can be exploited to regularise the inverse problem.
For some sparsifying basis $\bm{\Psi}$, we have $\bm{x} = \bm{\Psi\alpha}$, where $\bm{\alpha}$ is the set of coefficients representing $\bm{x}$ in the basis encoded in $\bm{\Psi}$.
The model parameters $\bm{m}$ sampled by MCMC are the coefficients $\bm{\alpha}$, and as such the forward operator is given by
\begin{equation}
    \bm{G}(\bm{\alpha}) = \bm{\Phi}(\bm{\Psi\alpha}).
    \label{eqn:forward}
\end{equation}
For non-linear inverse problems, i.e.\ where $\bm{\Phi}$ is a non-linear measurement operator, a common choice is to linearise via a Taylor expansion around an initial guess model.
While this usually leads to acceptable results in weakly non-linear problems, linearisation errors and the need for a good initial guess model can lead to artefacts in the recovered model.
If the forward problem (equation~\ref{eqn:forward}) is computationally fast, sampling methods are typically best for strongly non-linear problems.

\subsection{Harmonic representations of spherical images}
We consider functions $f\thetaphi$ defined on the sphere \STWO{} with colatitude $\theta \in [0, \pi]$ and longitude $\phi \in [0, 2\pi)$.
The canonical orthogonal basis functions for \STWO{} are the spherical harmonics 
\begin{equation}
   Y_{lm}\thetaphi = \sqrt{\frac{2\ell + 1}{4\pi}\frac{(\ell - m)!}{(\ell + m)!}}P_{\ell}^m(\cos\theta)e^{im\phi}  
\end{equation}
for non-negative integers $\ell$ and integers $m \leq |\ell|$, denoting angular degree and angular order, respectively.
As per the Condon-Shortley phase convention, the associated Legendre functions $P_{\ell}^m(x)$ include a $(-1)^m$ phase factor, ensuring the conjugate symmetry relation
\begin{equation}
    Y_{\ell,-m}\thetaphi = (-1)^m Y_{\ell m}^*\thetaphi,
\end{equation} 
where $^*$ denotes complex conjugation.
Any square integrable scalar function on \STWO{} can be represented in the frequency domain by its harmonic coefficients $f_{lm} \in \mathbb{C}$, obtained by projecting $f$ onto the spherical harmonic basis functions using the inner product (forward spherical harmonic transform)
\begin{equation}
    f_{lm} = \langle f,Y_{\ell m} \rangle = \int_{\STWO}f\thetaphi Y_{\ell m}^*\thetaphi \sin\theta d\theta d\phi.
    \label{eqn:sht}
\end{equation}
The function $f(\theta,\phi)$ can be recovered exactly from its spherical harmonic coefficients (inverse spherical harmonic transform) by
\begin{equation}
    f\thetaphi = \sum_{\ell=0}^{\infty}\sum_{m=-\ell}^{\ell} f_{\ell m}Y_{\ell m}\thetaphi.
    \label{eqn:isht}
\end{equation}
This formalism can be generalised to spin-$s$ fields \citep{Newman1966,Goldberg1967}, common to astrophysical spherical signals \citep{McEwen2015,Wallis2021}, which are characterised by local rotations $\chi \in [0, 2\pi)$ in the tangent plane at a point \thetaphi,
\begin{equation}
    _sf'\thetaphi = e^{-is\chi}\ _sf\thetaphi
\end{equation}
The spin-weight $s$ of a function can be increased or decreased by applying the spin-raising or spin-lowering operators, $\eth, \overline{\eth}$, respectively.
\begin{equation}
    \eth \equiv - \sin^s\theta\left(\frac{\partial}{\partial\theta} + \frac{i}{\sin\theta}\frac{\partial}{\partial\phi}\right)\sin^{-s}\theta,
\end{equation}
\begin{equation}
    \overline{\eth} \equiv - \sin^{-s}\theta\left(\frac{\partial}{\partial\theta} - \frac{i}{\sin\theta}\frac{\partial}{\partial\phi}\right)\sin^s\theta.
\end{equation}
Note how these operators are defined based on the spin of the function to which they are applied.
Using these operators, one can obtain the spin-weighted spherical harmonics from the original scalar (spin-0) spherical harmonics, which form an orthogonal basis for spin functions on the sphere
\begin{equation}
    _sY_{\ell m}\thetaphi = \begin{cases}
        0, & \ell < |s| \\
        \sqrt{\frac{(\ell - s)!}{(\ell + s)!}}\eth^s Y_{\ell m}\thetaphi, & 0 \leq s \leq \ell \\
        \sqrt{\frac{(\ell + s)!}{(\ell - s)!}}(-1)^s\overline{\eth}^{-s} Y_{\ell m}\thetaphi, & -\ell \leq s \leq 0 \\
    \end{cases}
\end{equation}
As such, spin-$s$ fields can easily be decomposed and reconstructed in terms of spin spherical harmonics in a manner similar to equations~\ref{eqn:sht} and~\ref{eqn:isht}.

In this work we consider bandlimited signals, that is signals such that $_sf_{\ell m} = 0,\; \forall\; \ell\geq L$ for some bandlimit $L$.
We discretise spherical signals according to the McEwen-Wiaux (MW) sampling theorem \citep{McEwen2011}.
This equiangular sampling theorem has theoretically exact and efficient spherical harmonic transforms and requires fewer discrete points on the sphere than other sampling theorems \citep[e.g.][]{Driscoll1994,Gorski2005}.
It also exploits a relationship between spin spherical harmonics and Wigner functions \citep{Goldberg1967,McEwen2011} to avoid repeated applications of spin-raising/lowering operators for fast spin spherical harmonic transforms.

\subsection{Wavelet representations of spherical images}
For our sparsifying basis, we consider scale-discretised axisymmetric wavelets \citep{Wiaux2008,Leistedt2013,McEwen2018}, i.e., the wavelets are azimuthally symmetric when placed at the poles.
There exist directional wavelets which vary azimuthally \citep[e.g.][]{Wiaux2008,McEwen2015,McEwen2018} and may be desirable in certain applications.
However, such wavelets require transforms on the rotation group \sothree{} which are more computationally expensive than transforms on \STWO{} \citep{McEwen2015b}.
The spherical wavelet transform is defined as the convolution of $f$ and the wavelets $\Psi^j\thetaphi$.
Convolution on the sphere is defined as the inner product of $f$ with wavelets that have been rotated over the surface of the sphere by some operator $\mathcal{R}_{\thetaphi}$, analogous to standard convolution on the plane where the wavelet is laterally translated over the image.
As such, the wavelet coefficients $W^{\Psi^j}\in\STWO$ are given by
\begin{equation}
    W^{\Psi^j}\thetaphi = \langle f, \mathcal{R}_{\thetaphi}\Psi^j\rangle.
    \label{eqn:fswt}
\end{equation}
The wavelets $\Psi^j$ cover a range of scales $J_0\leq j \leq J$, which extract the highest frequency (high $\ell$ information of $f$).
The lowest frequency information (low $\ell$) is extracted by a scaling function $\Upsilon\thetaphi$ in a similar manner:
\begin{equation}
    W^{\Upsilon}\thetaphi = \langle f, \mathcal{R}_{\thetaphi}\Upsilon\rangle.
\end{equation}
The wavelets and scaling function are defined on the harmonic line such that the wavelets can be seen as narrow pass-band filters and the scaling function a low-pass filter that includes the zero frequency component.
For more details on choosing the width of the wavelets on the harmonic line we refer the interested reader to \citet{Leistedt2013}.
The wavelets and scaling function are defined such that they satisfy an admissibility condition which allows $f$ to be decomposed and recovered exactly from its wavelet coefficients.
The reconstruction is given by
\begin{equation}
    f\thetaphi = \sum_{\Gamma}\int_{\STWO} W^{\Gamma}\thetaprimephiprime(\mathcal{R}_{\thetaprimephiprime}\Gamma)\thetaphi \sin{\theta'}d\theta'd\phi'
    \label{eqn:iswt}
\end{equation} 
where $\Gamma \in \{\Upsilon,\Psi^j\}$.
Note that in practice the spherical wavelet transform is computed in harmonic space \citep{Leistedt2013}, meaning the forward and inverse wavelet transforms (equations~\ref{eqn:fswt}-\ref{eqn:iswt}) implicitly involve both a forward and an inverse spherical harmonic transform \citep{Wallis2017}, which dominate the computational complexity.
The spherical harmonic and axisymmetric wavelet transforms used in this work are implemented in the packages \href{https://github.com/astro-informatics/ssht}{\texttt{PYSSHT}}\footnote{\url{https://github.com/astro-informatics/ssht}} and \href{https://github.com/astro-informatics/s2let}{\texttt{PYS2LET}}\footnote{\url{https://github.com/astro-informatics/s2let}}, respectively.

\section{Framework for posterior sampling on the sphere}\label{sec:framework}
In this section we outline our framework for sampling the posterior of spherical inverse problems.
We give the details of the high-dimensional sampling algorithm we use and the modifications necessary for the sphere.
Extra consideration is given to the forward operator, and finally we define a measure of uncertainty that can be calculated from the posterior samples.
\subsection{Proximal MCMC}
\subsubsection{Moreau-Yosida envelope and proximal mapping}
The $\lambda$-Moreau-Yosida envelope of some concave function $h(\bm{x})$ is given by \citep{Moreau1962,Bauschke2017} 
\begin{equation}
    h^{\lambda}(\bm{x}) = \min_{\bm{u}}\{h(\bm{u}) + \|\bm{u} - \bm{x}\|^2_2/(2\lambda)\},
\end{equation}
where $\lambda > 0$.
This is a smooth approximation of $h$ with many desirable properties.
Firstly, $h^{\lambda}$ can be made arbitrarily close to $h$ by making $\lambda$ small.
Secondly, the minimisers of $h^{\lambda}$ are the same as the minimisers of $h$.
Thirdly, $h^{\lambda}$ is continuously differentiable even if $h$ is not.
The gradient of $h^{\lambda}$ is given by
\begin{equation}
    \nabla h^{\lambda}(\bm{x}) = [(\bm{x} - \mathrm{prox}^{\lambda}_{h}(\bm{x}))]/\lambda,
    \label{eqn:grad_MY}
\end{equation}
where $\mathrm{prox}^{\lambda}_{h}(\bm{x})$ is the proximal mapping of $h$ \citep{Moreau1962}
\begin{equation}
    \mathrm{prox}^{\lambda}_{h}(\bm{x}) = \argmin_{\bm{u}}\{h(\bm{u}) + \|\bm{u} - \bm{x}\|^2_2/(2\lambda)\}.
\end{equation}
Rewriting equation~\ref{eqn:grad_MY} as 
\begin{equation}
    \mathrm{prox}^{\lambda}_{h}(\bm{x}) = \bm{x} - \lambda\nabla h^{\lambda}(\bm{x})
\end{equation}
is reminiscent of the standard first-order finite differences approximation $h(x_0+a) \approx h(x_0) + ah'(x_0)$, which shows that $\mathrm{prox}^{\lambda}_{h}$ may be interpreted as a gradient step in $h^{\lambda}$ with step size $\lambda$.
The key property here is that the proximal map can be used to minimise $h^{\lambda}$ and, by extension, the non-differentiable $h$ since they share the same minimisers \citep{Parikh2014}.
For an excellent introduction to proximal mappings and algorithms, see \citet{Parikh2014} and \citet{Combettes2011}.

\subsubsection{Proximal Langevin algorithm}
The proximal MCMC method developed by \citet{Pereyra2016} is based on Langevin MCMC \citep{Roberts1996}, a gradient-based sampling method, which we describe here.
Consider a Langevin diffusion process $Y(t)$ for $0\leq t \leq T$ associated with a stationary pdf $\pi$.
This process is given by the stochastic differential equation 
\begin{equation}
    dY(t) = \frac{1}{2}\nabla\log{\pi[Y(t)]}dt + dW(t)
    \label{eqn:langevindiff}
\end{equation}
for some Brownian motion $W$.
We use the forward Euler discrete time approximation with step size $\delta$
\begin{equation}
    \bm{m}^{(i+1)} = \bm{m}^{(i)} + \frac{\delta}{2}\nabla\log{\pi[\bm{m}^{(i)}]} + \sqrt{\delta}\bm{w}^{(i)},
    \label{eqn:ULA}
\end{equation}
where $\bm{m}$ is the discretised Langevin diffusion and $\bm{w}^{(i)} \sim \mathcal{N}(0,\mathbb{I})$.
This is the Unadjusted Langevin Algorithm (ULA) \citep{Roberts1996}.
Under certain regularity conditions, ULA produces samples that converge to an ergodic measure close to $\pi$.
A MH acceptance step can be added to remove the approximation error, giving the Metropolis-Adjusted Langevin Algorithm (MALA) \citep{Roberts1996}.

A well known limitation of the Langevin and other gradient-based sampling algorithms is that they require differentiable pdfs, which is not the case for some popular priors such as the Laplace prior.
A proposed solution to this is to apply a Moreau-Yosida approximation to the non-differentiable terms in the posterior pdfs \citep{Pereyra2016,Cai2018I,Pereyra2020}.
The posterior pdf for our inverse problem (equation~\ref{eqn:bayes}) is of the form $\pi(\bm{m}) \propto \mathrm{exp}(-g(\bm{m}) -f(\bm{m}))$, where $g(\bm{m}) = \frac{1}{2\sigma^2}\|\bm{d}-\bm{G}(\bm{m})\|_2^2$ is our Gaussian data fidelity and $f(\bm{m}) = \mu\|\bm{m}\|_1$ is our non-differentiable Laplacian prior.
Applying a $\lambda$-Moreau-Yosida approximation to the prior, it then follows that the chain step for the Moreau-Yosida ULA (MYULA) is given by
\begin{eqnarray}
    \bm{m}^{(i+1)}&{}={}&\left(1-\frac{\delta}{\lambda}\right)\bm{m}^{(i)} + \frac{\delta}{\lambda}\mathrm{prox}_f^{\lambda}[\bm{m}^{(i)}]\nonumber\\
    &&{-}\:\delta\nabla g[\bm{m}^{(i)}] + \sqrt{\delta}\bm{w}^{(i)}.
    \label{eqn:myula}
\end{eqnarray}
The tuning parameter $\delta$ must be small for the forward Euler approximation (equation~\ref{eqn:ULA}) to converge, and \citet{Pereyra2016} argued that the optimal value for $\lambda$ is $\lambda = \delta / 2$.
We describe how to calculate the proximal mapping of our prior on the sphere (second term equation~\ref{eqn:myula}) in the following subsection.
If $\bm{G}$ is linear, the gradient of the data fidelity is straightforward to calculate using the adjoint
\begin{equation}
    \nabla g = \bm{G}^{\dagger}(\bm{Gm} - \bm{d})/\sigma^2.
    \label{eqn:gradg}
\end{equation}
It is straightforward to modify equation~\ref{eqn:gradg} if the data errors are characterised by a covariance matrix $\bm{C}$ rather than a single standard deviation $\sigma$.
It is important to note here that for spherical inverse problems, the forward operator $\bm{G}$ may include spherical harmonic transforms, and thus the adjoint transforms are also needed for equation~\ref{eqn:gradg}.
Further discussion about $\bm{G}$ in the spherical setting is given later in this section.
Algorithm~\ref{alg:myula} outlines the use of the MYULA chain in practice, highlighting the key steps and equations.
Again, a MH acceptance step can be added \citep{Pereyra2016,Cai2018I}, however in our experiments there was little improvement for the additional computational cost.

\begin{algorithm}
    \caption{MYULA on the sphere}\label{alg:myula}
    \begin{algorithmic}
    \STATE \textbf{INPUTS}: observed data $\bm{d}$, data errors $\sigma$, initial sample $\bm{m}^{(0)}$, $i=0$, $N$, $N_{\mathrm{thin}}$, $N_{\mathrm{burn}}$, quadrature weights $\bm{q}$, $\delta$, $\lambda$, $\mu$ 
    \STATE \textbf{OUTPUTS}: chain $\{\bm{m}^{(i)}:i=1,...,N\}$
    \WHILE{$i<N\times N_{\mathrm{thin}} + N_{\mathrm{burn}}$}
    \STATE Calculate gradient of data fidelity (eq~\ref{eqn:gradg})
    \STATE Calculate proximity map of prior (eq~\ref{eqn:soft})
    \STATE Calculate $\bm{m}^{(i+1)}$ (eq~\ref{eqn:myula})
    \IF{$i > N_{\mathrm{burn}}$}
    \IF{$\mathrm{mod}(i,N_{\mathrm{thin}}) = 0$}
    \STATE Save $\bm{m}^{(i+1)}$ to chain
    \ENDIF
    \ENDIF
    \STATE $i \mathrel{+}= 1$
    \ENDWHILE
    \end{algorithmic}
\end{algorithm}

\subsection{Modification for the sphere}
We sample a set of spherical wavelet coefficients that are defined at each point on the sphere (equation~\ref{eqn:fswt}).
The MW sampling theorem is equiangular, and as such we need to account for an overdensity of sampling points near the poles when we calculate the proximal mapping of our sparsity-promoting prior.
Proximal operators are generally calculated by a small convex optimisation problem \citep{Parikh2014}.
Fortunately, there exist closed-form representations for the proximal mapping of many common functions \citep{Combettes2011}, including the $\ell_1$-norm we use in our prior.
This is crucial for MCMC methods where the proximal mapping needs to be repeatedly computed.
It can be shown that for $f(\bm{m}) = \mu\|\bm{m}\|_1$,
\begin{equation}
    \mathrm{prox}^{\lambda}_f(\bm{m}) = \mathrm{soft}_{\lambda\mu}(\bm{m})
    \label{eqn:proxL1}
\end{equation}
where $\mathrm{soft}_{\lambda\mu}$ is the \textit{soft thresholding} operator with threshold ${\lambda\mu}$ defined as \citep{Combettes2011}
\begin{equation}
    \mathrm{soft}_{\lambda\mu}(m_i) = \begin{cases}
        0 & \mathrm{if }\;\; m_i \leq {\lambda\mu}, \\
        m_i - {\lambda\mu}\mathrm{sgn}(m_i) & \mathrm{if }\;\; m_i > {\lambda\mu}
    \end{cases}
    \label{eqn:soft}
\end{equation}
where $\mathrm{sgn}(x)$ is the sign of $x$, and can be applied component-wise to the components $m_i$ of $\bm{m}$.
To account for the spherical sampling overdensity, a weighting must be applied to parameters $\bm{m}$, which takes the form of a diagonal matrix $\bm{W}$ of quadrature weights \citep{McEwen2011} that vary with colatitude and bandlimit.
This can be seen as applying a component-wise prior on each of the elements of $\bm{m}$.
The proximal mapping changes only in that the threshold in equation~\ref{eqn:soft} becomes $\lambda\mu w_i$, where $w_i$ is the $i^{\mathrm{th}}$ diagonal element of $\bm{W}$.

\subsection{The forward operator}
A key consideration for any MCMC sampling algorithm is the speed of the forward modelling step, i.e.\ making predictions of the data $\bm{d}$ given a MCMC sample $\bm{m}$.
This is particularly important for spherical problems where the forward operator $\bm{G}$ may contain spherical harmonic transforms, which are, unfortunately, slow and typically scale in complexity as $\mathcal{O}(L^3)$ \citep{McEwen2011}.
As discussed in Section~\ref{sec:background}, the spherical harmonic transforms are implicit in the spherical wavelet transforms which form our sparsifying basis.
Expanding equation~\ref{eqn:forward} to see this, to sample the spherical wavelet coefficients in pixel space $\bm{\alpha}$ we have in practice
\begin{equation}
    \bm{G}(\bm{\alpha}) = \bm{\Phi \mathrm{S}}^{-1}\bm{\mathrm{WS}\alpha},
    \label{eqn:fullforward}
\end{equation}
where $\bm{\mathrm{S}}$ and $\bm{\mathrm{S}}^{-1}$ are the forward and inverse spherical harmonic transforms (equations~\ref{eqn:sht} and~\ref{eqn:isht}), respectively, and $\bm{\mathrm{W}}$ is the spherical wavelet transform in harmonic space \citep{Leistedt2013,Wallis2017}.
This is further complicated by the need for the adjoints of these transforms (equation~\ref{eqn:gradg}).
These four spherical harmonic transforms at each MCMC step immediately limits the bandlimits $L$ for which sampling the posterior may be feasible.
For reference, in our experiments we performed dummy inversions at $L=64$ and $L=128$ with an identity measurement operator ($\bm{\Phi}=\bm{I}$) which took around 3 and 17 days, respectively, for $10^6$ samples on a 2.5 GHz Intel Xeon Platinum 8180M processor.
While this may not be an issue for applications where information exists at relatively low degrees, for example in seismic tomography where maximum bandlimits are typically around $L=40$ \citep[e.g.][]{Ritsema2011,Chang2015}, it will be impractical for applications where the bandlimits of interest are higher \citep[e.g.][]{Price2020}.
To avoid spherical harmonic transforms in equation~\ref{eqn:fullforward}, one could instead sample the spherical harmonic coefficients of $\bm{\alpha}$, provided one can reformulate the measurement operator $\bm{\Phi}$ appropriately.
We note however that it is possible that, as shown in the following section, that the harmonic measurement operator may be slower than the pixel space operator even with spherical wavelet transforms.
Further, it is conceivable that, depending on how the measurement operator scales, the choice of harmonic or pixel space will depend on the bandlimit $L$.
In either case, at least one spherical harmonic transform will be needed for the prior, as there is no reason to expect the harmonic representation of the wavelet coefficients to also be sparse.
As such, spherical harmonic transforms are unavoidable, and it is crucial that these are computed as efficiently as possible, for example via repeated exploitation of fast Fourier transforms as in \citet{McEwen2011}.
This all highlights the special consideration that must be given to the forward operator for spherical inverse problems.

\subsection{Spherical wavelet parameterisation}
As discussed in Section~\ref{sec:background}, our inverse problem is parameterised using axisymmetric spherical wavelets \citep{Wiaux2008,Leistedt2013,McEwen2015} to promote sparsity.
We can exploit the multiresolution character of these wavelets for further computational savings.
By construction, the wavelets at each scale $j$ each have different bandlimits $k_j=B^{j+1}\leq L$, where $B$ is a wavelet scale parameter \citep{Leistedt2013}.
By using a sampling theorem, the transforms at each spherical wavelet scale can be performed up to their own bandlimit $k_j$, and only the minimum number of samples on the sphere at that bandlimit are needed.
This multiresolution transform gives a four to five times speed up of the spherical wavelet transforms \citep{Leistedt2013}, and also dramatically reduces the dimensionality of our parameter space, compared to a full resolution transform where each wavelet scale is sampled at the overall bandlimit $L$.
As an example, for parameters $L=32,\;B=2,\;J_0=2$, the full resolution algorithm has \num{10080} wavelet coefficients, compared to only \num{4676} for the multiresolution algorithm.
While a full assessment of how this affects the convergence speed of the MCMC chain in terms of number of required steps is beyond the scope of this paper, we found real-time speed-ups and significant memory savings in our experiments.

\subsection{Uncertainty quantification}
By collecting samples from the posterior we can calculate any measure of uncertainty.
For example, a common choice in Bayesian statistics is the credible intervals $[\xi_i^{-}, \xi_i^{+}]$ of the model parameters.
These intervals contain the values that can be taken by parameter $m_i$ with probability $(1 - \alpha)$, for some chosen small $\alpha$
\begin{equation}
    p(m_i \in [\xi_i^{-}, \xi_i^{+}]|\bm{d}) = 1 - \alpha
    \label{eqn:ci_range}
\end{equation}
The lower and upper interval limits are calculated as the $\frac{\alpha}{2}$ and $1-\frac{\alpha}{2}$ quantiles, respectively, of the posterior.
We note that having sampled the spherical wavelet coefficients, inverse spherical wavelet transforms will be necessary to obtain an uncertainty map in real space as opposed to wavelet space.
This can be expensive for the same reasons as previously discussed with respect to the forward operator, although to a much lesser extent after burn-in and thinned samples have been discarded.
Of course if the desired summary statistic is linear in the model parameters (e.g.\ mean) then this can be calculated in wavelet space and only requires a single spherical wavelet transform.
Importantly, in this way we get uncertainties at the pixel level when viewed in real space.
As previously discussed, current uncertainty quantification methods in similar contexts only work on superpixels \citep{Price2019}.

\section{Seismological surface wave phase velocity maps}\label{sec:GST}
The deep structures of the Earth's interior are best illuminated from seismic waves.
Waves generated from earthquakes travel through the Earth, interacting with the materials along the way.
Inverting the measurements of these waves recorded at seismic receivers, typically located at the Earth's surface, reveals the layers and structures in the Earth \citep{Rawlinson2010}.
Depending on the type and frequency of the waves, this can be done to image a range of scales from local applications (e.g.\ imaging oil, gas, geothermal fields) to global problems (e.g.\ imaging the whole Earth's mantle, down to ${\sim}$\SI{2800}{\kilo\metre} depth).
The seismic imaging the Earth's interior in three dimensions is known as seismic tomography.

In this section we introduce the common problem in global seismic tomography of building surface wave phase velocity maps, which we use as an illustrative example.
Being a 2-D problem that can be described by relatively simple theory, this is a natural example of application that is well suited to illustrate our framework for sampling the posterior for spherical inverse problems.
The resolution requirements in global seismic tomography are also much lower than for full-sky mass-mapping, as discussed in the next section, typically only requiring $L\leq40$.
This makes our method directly applicable to this problem.

\subsection{Surface wave phase velocity maps}
Mapping the phase velocity of surface waves is a common problem in seismic tomography \citep{Trampert1995,Ekstrom1997,Ekstrom2011}.
Seismic surface waves generated from earthquakes travel along the Earth's surface.
These waves are dispersive in that their velocity depends on the wave period, with different wave periods being sensitive to the structures at different depths in the Earth's interior \citep{Dahlen1998}.
Phase velocity maps show how the velocity of surface waves at a given period varies due to lateral heterogeneities in the Earth's composition and temperature.  
Hence, creating these maps for waves of different periods is often a first step towards building 3D models of the Earth's mantle \citep[e.g.][]{Durand2015}.

For a particular wave period, one can measure from a seismogram the average phase velocity along the path that the wave has travelled between a seismic source and receiver.
Inverting measurements from many crossing paths, that would ideally cover the Earth uniformly, produces a phase velocity map.
In practice, the distribution of paths is determined by the locations of earthquakes, typically along tectonic plate boundaries, and the locations of seismic stations predominantly in the continents of the northern hemisphere, making this an ill-posed inverse problem.
On global scales, long-period ($T > 25$ s) phase velocity measurements are typically modelled using linearised ray theory (an infinite frequency approximation analogous to geometrical optics), also known as the great-circle approximation \citep[e.g.][]{Woodhouse1984,Parisi2016} and inverted using least-squares algorithms \citep[e.g.][]{Tarantola1982,Trampert1995,Ekstrom2011,Durand2015}.
In this framework, the path travelled by the seismic wave is assumed to correspond to the great circle between the source and the receiver.
The observed mean phase velocity anomaly $\langle{\delta c/c_0}\rangle$  for a given source-receiver pair is given by the average of the phase velocity field along the minor arc great circle $S$,
\begin{equation}
    \left\langle{\frac{\delta c}{c_0}}\right\rangle = \frac{1}{\Delta}\int_S\frac{\delta c}{c_0}(\theta, \phi) dS \, ,
    \label{eqn:pathint}
\end{equation}
where $c_0$ is the phase velocity value computed for a reference Earth model, $\delta c = c - c_0$, with $c$ being the actual phase velocity, and $\Delta$ is the length of path $S$.

Although there exist more complete theories describing phase velocity anomalies, notably including non-linear and finite frequency effects, recent studies have shown that the great-circle approximation accurately predicts the phase of long-period fundamental mode surface waves for current global tomography models \citep[e.g.][]{Parisi2016,Godfrey2019}.
Hence, in this study we focus on the great-circle approximation.
For simplicity, we also only invert for isotropic lateral variations in phase velocity.
We note, however, that our proximal MCMC approach can in principle handle these additional physical parameters (e.g.\ anisotropy, the directional dependence of wave velocity) with some slight modification.
\subsection{Pixel space path integration}
\begin{figure}
    \includegraphics[width=0.45\textwidth]{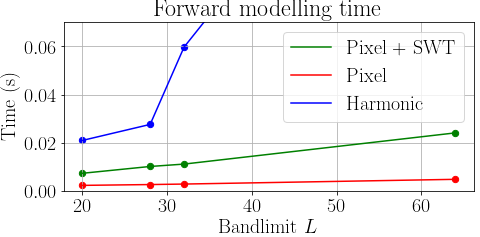}
    \caption{Time taken to perform the forward modelling in pixel space (red), in pixel space with a spherical wavelet transform (green), and in harmonic space (blue).
    The point for the harmonic approach at bandlimit $L=64$ is well beyond the vertical scale at around 0.2 s.
    Timings performed on a 2020 MacBook Pro with an Apple M1 processor.}
    \label{fig:harmvpix_time}
\end{figure}

The common way to compute path integrals on the sphere (equation~\ref{eqn:pathint}) is to rotate the coordinate system such that the path lies along the equator, which is easily expressed in terms of Wigner-D matrices and the spherical harmonic coefficients of the spherical signal \citep{Dahlen1998}.
For a dataset with $N_{\mathrm{paths}}$ paths, the measurement operator can be represented by a dense matrix $\bm{\Phi}_h \in \mathbb{C}^{ N_{\mathrm{paths}} \times L^2}$ acting on the spherical harmonic coefficients $\hat{\bm{x}} \in \mathbb{C}^{L^2}$.
For seismic datasets typically consisting of $\mathcal{O}(10^5)$ paths, this dense matrix multiplication can be quite slow.
Our approach is to instead measure the path integral directly on the pixelised sphere using a sparse matrix $\bm{\Phi}_p \in \mathbb{R}^{N_{\mathrm{paths}}\times N_{\mathrm{pixels}}}$, where each element of $\bm{\Phi}_p$ is a weight representing the normalised distance each path travels in a pixel.
This effectively approximates the integral as a weighted Riemann sum over the pixelised function $\bm{x} \in \mathbb{R}^{N_{\mathrm{pixels}}}$.
The adjoint operator is trivially the transpose of $\bm{\Phi}_p$.
The first step for our numerical path integration is to find discrete geographical points along the great circle minor arc between a source and a receiver using spherical trigonometry.
For the discretisation, a sampling rate of about 200 points per radian (3.5 points per degree) was generally sufficient for this work.
Each of the geographical points along the path is then mapped to its nearest MW sampling point.
This mapping assigns a weight to each MW sampling point, which is given by
\begin{equation*}
    w_{tp} = \frac{n_{tp}}{s\Delta},
\end{equation*}
where $n_{tp}$ is the number of geographical points on the path that are closest to MW sampling point indexed in the $\theta$ and $\phi$ directions by $t$ and $p$, respectively, $s$ is the path sampling rate and $\Delta$ is the full path length.
This can easily be done for each path of the dataset in parallel to build the full measurement operator.

\begin{table}
    \renewcommand{\arraystretch}{1.3}
    \caption{Accuracy of pixel space path integration}
    \label{tab:harmvpix}
    \centering
    \begin{tabular}{ccc}
        \textbf{Bandlimit} $L$ & \textbf{Mean Diff.} (\%) & \textbf{R2E} \\
        \hline
        20 & -0.01 & \num{2.14e-4} \\
        28 & -0.02 & \num{1.52e-4} \\
        32 & -0.02 & \num{1.34e-4} \\
        64 & -0.01 & \num{5.64e-5} \\
        \hline
    \end{tabular}
\end{table}

Figure~\ref{fig:harmvpix_time} compares the forward modelling time for both the harmonic and pixel space path integrations, $\bm{\Phi}_h$ and $\bm{\Phi}_p$, respectively, for bandlimits $L \in \{20, 28, 32, 64\}$ and a realistic set of ray paths (see Figure~\ref{fig:synth_uncert} bottom).
Also shown are the times for the pixel space integration with an additional spherical wavelet transform $\bm{\Phi}_p\bm{\Psi}$, as required when sampling wavelet coefficients (equation~\ref{eqn:forward}) instead of sampling the image directly.
Clearly integration in pixel space is much faster than in harmonic space, even with the computational overhead of the spherical wavelet transforms.
Pixel space integration also scales better to higher bandlimits.
This is due to the extreme sparsity (less than 2\% nonzero elements) of $\bm{\Phi}_p$, whereas $\bm{\Phi}_h$ is generally dense.
Table~\ref{tab:harmvpix} shows the mean percentage difference and the relative squared error $\mathrm{R2E}=\|\bm{d}_{\mathrm{harm}} - \bm{d}_{\mathrm{pix}}\|_2^2 / \|\bm{d}_{\mathrm{harm}}\|_2^2$ between predictions made in harmonic space, $\bm{d}_{\mathrm{harm}}$, and in pixel space, $\bm{d}_{\mathrm{pix}}$, when performed on the ground truth map we use in our synthetic experiment (see Figure~\ref{fig:synth} top).
As can be expected, as the bandlimit increases the error in the pixel space integration decreases.
Crucially, even at relatively low bandlimits the pixel space integration is sufficiently accurate.

\subsection{Results and Discussion}
\label{sec:results}
In this section we present the results of a synthetic test and real data inversions.
We use MYULA to sample the axisymmetric wavelet coefficients $\bm{\alpha}$ of the spherical image $\bm{x}$, using the sparse measurement operator described in the previous section.

\subsubsection{Synthetic experiment}

\begin{figure}
    \includegraphics[width=0.48\textwidth]{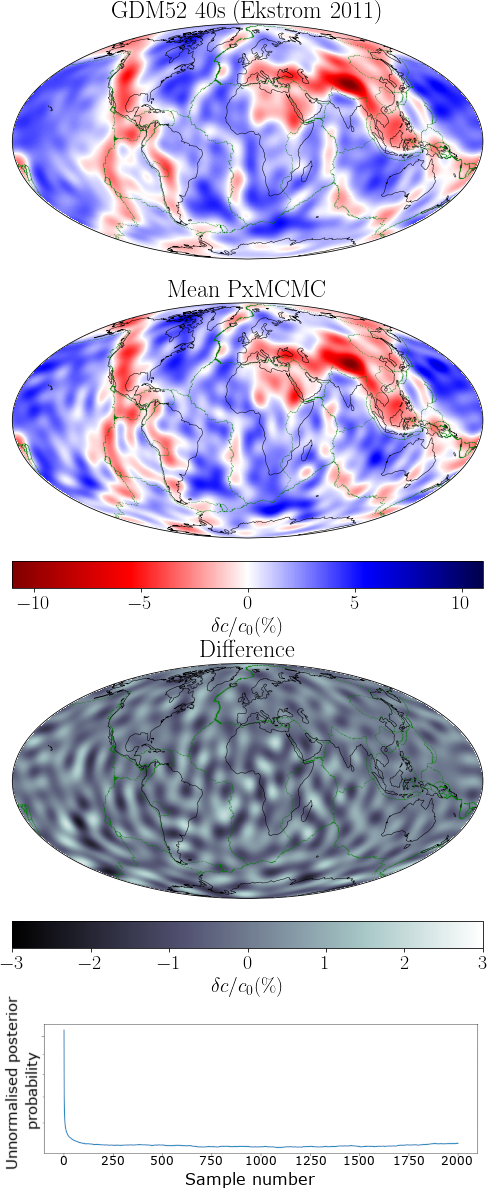}
    \caption{Synthetic GDM52 recovery experiment.
    (top) Ground truth from \protect\citet{Ekstrom2011}.
    (Middle) Mean solution from proximal MCMC\@.
    (Bottom) difference between the truth and our solution.
    All the maps show perturbations in phase velocity ($\delta c/c_0$) with respect to the reference model PREM \protect\citep{Dziewonski1981}.
    Green lines show the tectonic plate boundaries \protect\citep{Bird2003}.
    The bottom panel shows the unnormalised posterior probability throughout the MCMC chain, indicating that the sampling has converged.}
    \label{fig:synth}
\end{figure}

\begin{figure}
    \centering
    \includegraphics[width=0.5\textwidth]{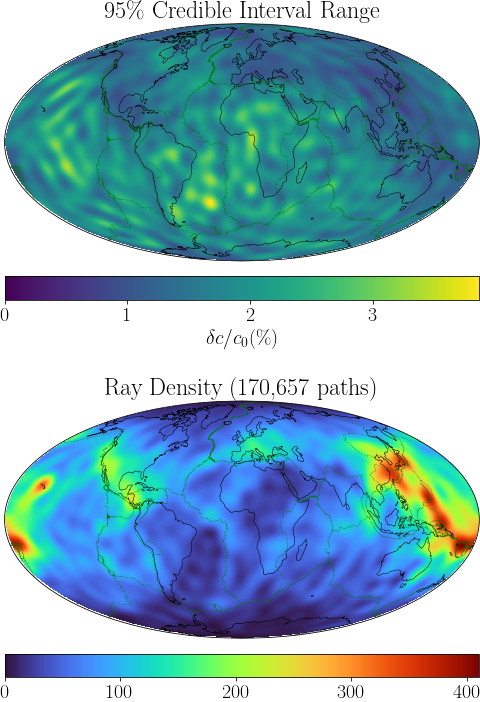}
    \caption{Synthetic GDM52 recovery experiment.
    (top) 95\% credible interval range (equation~\ref{eqn:ci_range}) calculated on our MCMC samples in image space ($\bm{x}$) in units of percentage deviation from the reference value, as in Figure~\ref{fig:synth}.
    (bottom) Row sum of path matrix $\bm{\Phi}$ (unitless) representing the density of rays in the dataset.}
    \label{fig:synth_uncert}
\end{figure}

As a synthetic example, we use the global phase velocity model GDM52 \citep{Ekstrom2011} at a wave period of $T=40$ s as a ground truth $\bm{x}$.
Surface waves of this period are mainly sensitive to Earth structure at depths of around 100 km \citep{Dahlen1998,Durand2015}, so the image (Figure~\ref{fig:synth}) shows well-known tectonic features such as slow anomalies along spreading ridges \citep[e.g.][]{Ekstrom2011}.
The model is bandlimited to $L=28$, corresponding to \num{3724} wavelet coefficients using the multiresolution wavelet transform.
We create a synthetic dataset $\bm{d}$ 
\begin{equation}
    \bm{d} = \bm{\Phi x} + \bm{n},
\end{equation}
where $\bm{\Phi}$ is our pixel-space forward operator and $\bm{n}\sim\mathcal{N}(0,\sigma)$, where $\sigma$ is the standard deviation of the predictions $\bm{\Phi x}$ to simulate noise in observed data.
$\bm{\Phi}$ is constructed using the same paths as those used to originally build GDM52 \citep{Ekstrom1997,Ekstrom2011}, thereby ensuring a realistic and non-uniform spatial distribution of the data.
In this case we have \num{179657} paths.
We use the signal-to-noise ratio
\begin{equation}
    \mathrm{SNR}(\bm{x}_0) = 20\log_{10}\left(\frac{\|\bm{x}\|_2}{\|\bm{x} - \bm{x}_0\|_2}\right)
\end{equation}
and relative squared error
\begin{equation}
    \mathrm{R2E}(\bm{x}_0) = \frac{\|\bm{d} - \bm{\Phi \bm{x}}_0\|_2^2}{\|\bm{d}\|_2^2}
\end{equation}
to assess the reconstruction accuracy and predictive accuracy, respectively, of our chosen point solution $\bm{x}_0$, which in this case we choose to be the post-burn mean mean of our MCMC samples.
We perform $10^6$ MCMC steps, saving every 500\textsuperscript{th} sample.
The first half of the saved samples are discarded as a burn-in when calculating our mean solution and uncertainty.
This takes the sampling well beyond the point of convergence, which we take to be the point where there is no longer a significant change in posterior probability (see Figure~\ref{fig:synth}).
This inversion takes about 20 hours on a 2.5 GHz Intel Xeon Platinum 8180M processor.
The tuning parameters are set to $\mu=500$ and $\delta=10^{-6}$.
Figure~\ref{fig:synth} shows the ground truth, the post-burn mean of our proximal MCMC samples and the difference between the two.
Our solution has an excellent data fit ($\mathrm{R2E}=$ \num{9.96e-3}) and model recovery ($\mathrm{SNR}=8.81$ dB).
Differences between the ground truth and our mean solution are small on average (0.5\%) with some small-scale blobs of larger differences.
The majority of these blobs occur in the southern hemisphere where data coverage is poorer (see Figure~\ref{fig:synth_uncert}).

\begin{figure*}
    \includegraphics[width=\textwidth]{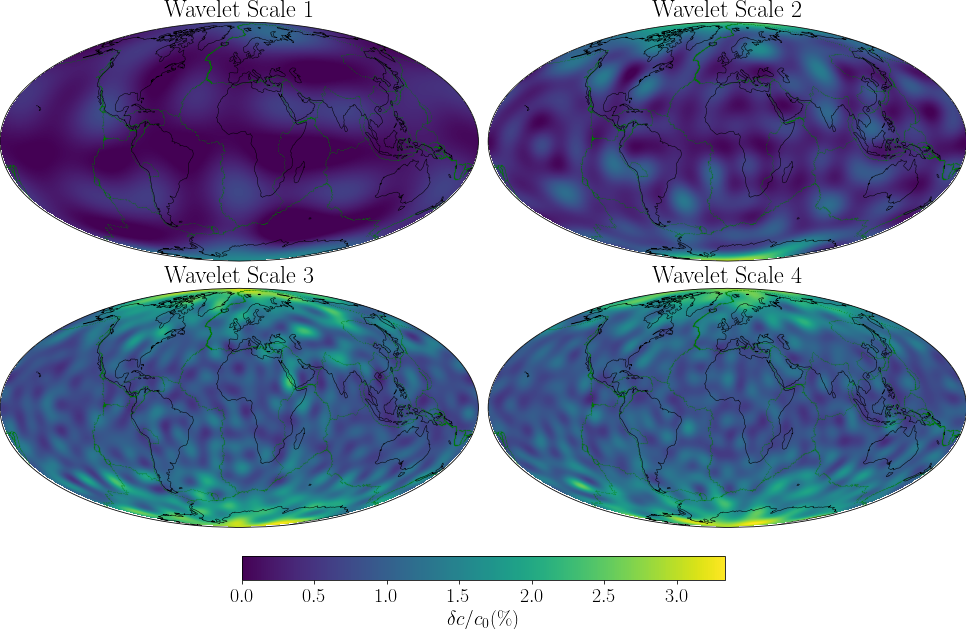}
    \caption{Synthetic GDM52 recovery experiment.
    95\% credible interval ranges of spherical wavelet coefficients at different scales.
    The range of the colourbar is the same for all maps in units of percentage deviation from the reference value, as in Figure~\ref{fig:synth} and~\ref{fig:synth_uncert}.}
    \label{fig:wav_ci}
\end{figure*}

We show the map of 95\% credible interval ranges as well as a map showing the density of ray paths of our data set in Figure~\ref{fig:synth_uncert}.
Here the uncertainty is calculated on the image space representation of our solution (i.e.\ $\bm{x}$).
As can be expected, on the whole we find lower uncertainties where we have a higher density of data in areas such as eastern Asia and western US.
We also see much smaller scale regions of higher uncertainty.
Looking at the uncertainties in wavelet space (i.e.\ $\bm{\alpha}$, which is sampled by the proximal MCMC) in Figure~\ref{fig:wav_ci} it is clear that the smaller scale wavelet coefficients have higher uncertainty.
Thus, the patterns of differences between our chosen point solution and the truth (Figure~\ref{fig:synth}) are captured by the uncertainty of our sampled parameters.

\subsubsection{Real Data Inversions}
\begin{figure*}
    \centering
    \includegraphics[width=\textwidth]{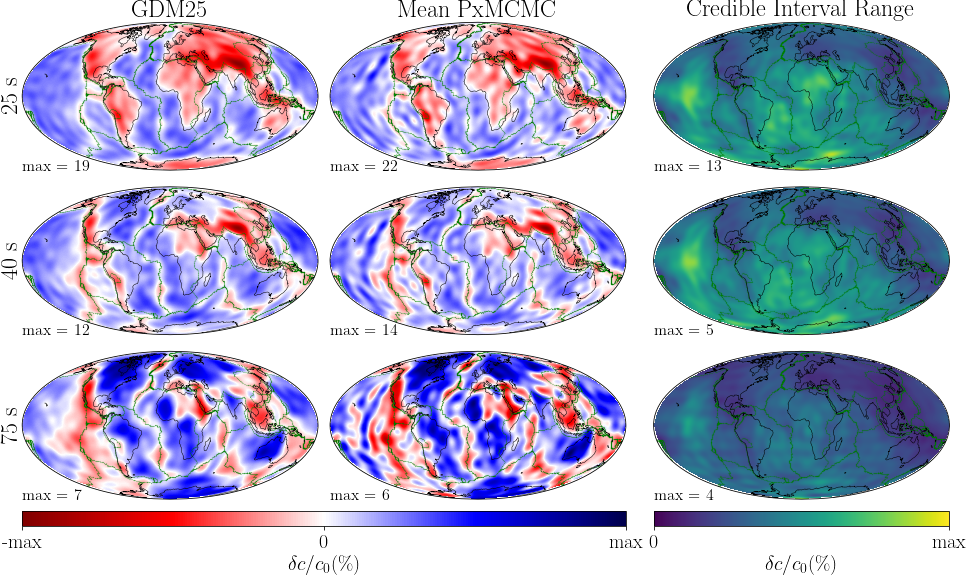}
    \caption{Real data inversions.
    Phase velocity maps from GDM52 \protect\citep{Ekstrom2011} (left), mean proximal MCMC solutions from this study (middle) and credible interval ranges (right) for wave periods 25, 40 and 75 s (top to bottom).
    All the maps show perturbations in phase velocity ($\delta c/c_0$) with respect to the reference model PREM \protect\citep{Dziewonski1981}.}
    \label{fig:real_data_results}
\end{figure*}

To demonstrate our method on real data, we invert the same data that was used to build GDM52 \citep{Ekstrom1997,Ekstrom2011} at wave periods $T = 25,\; 40,\; 75$ s, which have strong sensitivity to about 100 km, 160 km and 300 km depths, respectively \citep{Dahlen1998,Durand2015}.
The results of these inversions are shown in Figure~\ref{fig:real_data_results}.
The datasets at these wave periods consist of about \num{103633}, \num{179657} and \num{286302} paths, respectively.
Again we use a bandlimit of $L=28$.
For these inversions, we perform \num{750000} chain steps, saving every $500^{\mathrm{th}}$ sample and discarding the first 500 saved samples as burn-in.
This takes between 8 and 33 hours, depending on the number of ray paths, on a 2.5 GHz Intel Xeon Platinum 8180M processor.
The tuning parameter $\delta$ is chosen on a case-by-case basis.
Our mean solutions show all the expected velocity anomalies, being very similar to the GDM52 phase velocity maps (Figure~\ref{fig:real_data_results}, left).
For example, the $T = 25$ s map (Figure~\ref{fig:real_data_results}, top) shows a clear distinction between the slow continents and the fast oceans.
On the other hand, the $T = 40$ s map (Figure~\ref{fig:real_data_results}, middle) depicts a good correlation between slow anomalies and plate boundaries, while the $T = 75$ s map (Figure~\ref{fig:real_data_results}, bottom) shows deeper mantle signals, such as high velocities associated with cratons.
A notable difference with GDM52 (Figure~\ref{fig:real_data_results}, left) is a north-south streak of fast velocities off the coast of the western US seen at all wave periods.
This is a well-known artefact resulting from not modelling azimuthal anisotropy \citep{Forsyth1975,Ekstrom2011}.
This streak corresponds to a region of high uncertainty in our solutions (Figure~\ref{fig:real_data_results}, right).
Again, the uncertainty maps correlate with ray density as expected.

We emphasise that the main purpose of this study is not to build improved phase velocity maps, which can be constructed quickly using, e.g.
least squares approaches \citep{Tarantola1982,Trampert1995}, but rather to illustrate and validate our framework for sampling the posterior of spherical inverse problems with a useful, well-known first application.
Hence, we do not consider more sophisticated theoretical formulations, such as, e.g.\ full ray theory \citep{Ferreira2007}, finite frequency theory \citep{Zhou2005}, including anisotropic effects \citep{Ekstrom2011}, etc.
Future work will focus on the application of the method to more sophisticated problems, such as e.g.\ depth inversions using non-linear theory.

\section{Low-resolution Cosmological Mass-Mapping}
\label{sec:CMM}
One of the predictions of Einstein's theory of general relativity was that the gravitational influence of massive objects will cause light to bend around them \citep{Einstein1905a}.
As a result, distant objects often look distorted when observed by astronomers.
This is the phenomenon of gravitational lensing.
Light from distant stars and galaxies travels through the universe, bending around all the mass along the way to Earth.
Inverting measurements of the distortions in the images of the light source reveals the hidden masses that the light passed by \citep[e.g.][]{Kaiser1993}, including dark matter \citep{Heavens2009}, producing so-called cosmological mass maps.

In this section we demonstrate our framework on the retrieval of cosmological mass-maps from simulated data.
This is a 2-D problem described by a simple linear forward model.
We treat this as a simple demonstration and perform the inversion at a much lower bandlimit than would normally be required for cosmological mass-mapping.
Due to the poor scaling of the spherical wavelet transforms with $L$, performing the millions of transforms required at $L\sim\mathcal{O}(10^3)$ \citep{Wallis2021,Price2020} would simply be too slow.
However, with future computational improvements to speed up the transforms (e.g.\ by exploiting GPUs), our framework will be well-suited for this particular problem.

\subsection{Mass-mapping on the celestial sphere}
Images of galaxies are typically distorted, as the light they emit gets lensed by the  mass between us and the source galaxies.
Gravitational lensing occurs regardless of the nature of the intervening mass, and as such lensing is an excellent probe for dark matter \citep{Heavens2009}.
Mass-mapping maps the total density perturbation along a line of sight between a source galaxy and the observer based on measurements of the distortion of galaxy images.
Up until recently, lensing surveys only covered relatively small sky-fractions, so planar approximations were made.
As the area of coverage has increased with newer surveys, planar approximations are no longer valid \citep{Wallis2021}, resulting in mass-maps now being constructed on the sphere \citep[e.g.][]{DESY3,Price2020,Wallis2021}.

Gravitational lensing studies deal with two main fields: the convergence field $_0\kappa\thetaphi$, which causes magnification of the galaxy image; and the shear field $_2\gamma\thetaphi$, which causes rotation and stretching of the galaxy image \citep{Dodelson2017}.
Mass maps reveal the convergence field $_0\kappa\thetaphi$, which can be shown to be the integrated mass density along the line of sight \citep{Bartelmann2001}.
This is linearly related in spherical harmonic space to the spin-2 shear field $_2\gamma\thetaphi$, measured from observations of galaxy shapes, by a linear kernel $\mathcal{K}_{\ell}$ given by \citep{Kaiser1993,Wallis2021}
\begin{equation}
   \mathcal{K}_{\ell} = \frac{-1}{\ell(\ell + 1)}\sqrt{\frac{(\ell + 2)!}{{(\ell - 2)!}}}.
\end{equation}
Thus, our forward model in this case is given by $\bm{\Phi} = \bm{M}\;_2\bm{S}^{-1}\bm{K}\;_0\bm{S}$, where $\bm{K}$ encodes the linear kernel above and $\bm{M}$ is a masking matrix to account for areas on the sky without reliable data (e.g.\ the galactic plane and ecliptic).
$_2\bm{S}$ and $_0\bm{S}$ denote the spin-2 and spin-0 spherical harmonic transforms needed for to account for the spin symmetries of the shear and convergence fields, respectively.
From hereon in we drop the spin subscripts for clarity.
We note that in reality there exists a degeneracy between $\gamma$ and $\kappa$, and as a result the true observable is not the shear but the reduced shear $g = \gamma / (1-\kappa)$.
In the weak lensing regime, this non-linear effect is very small, so we ignore it here.
Accounting for non-linearities in our MCMC is possible, provided the relevant gradients of the non-linear forward model can be computed efficiently.

The choice of using sparsity promoting priors in this case is motivated by the need to recover non-Gaussian structures, particularly at high $\ell$.
These are created by non-linear structure growth of the density field throughout the evolution of the Universe.
Promoting sparsity has been used previously \citep[e.g.][]{Leonard2014,Lanusse2016,Price2020,Starck2021} to incorporate non-Gaussian structures, as have Wiener filters on a Gaussian prior \citep[e.g.][]{Jeffrey2018},log-normal priors \citep[e.g.][]{Bohm2017,Fiedorowicz2022} and physically informed power spectrum priors \citep[e.g.][]{Porqueres2021}.

A Bayesian sampling method for mass-mapping was recently implemented by \citet{Fiedorowicz2022}.
They used HMC to efficiently navigate the large parameter space with a log-normal prior.
In comparison with our framework, the log-normal prior is differentiable, which allows the gradient-based HMC to be used, with the exploitation of efficient autodifferentiation making this computationally tractable.
However, the log-normal prior makes certain assumptions about the cosmological parameters, and thus the mass maps produced by their method cannot be used for inference about cosmology, without additional further work \citep{Fiedorowicz2022}.
Our framework makes no such assumptions.

\subsection{Results and Discussion}
As a synthetic example, we attempt to reconstruct a simulated mass map from the Takahashi N-body simulation \citep{Takahashi2017}\footnote{Data available at \url{ http://cosmo.phys.hirosaki- u.ac.jp/takahasi /allsky raytracing/}.}.
Slices are provided at a range of redshifts, and we select redshift slice 16 corresponding to $z\sim1$.
This slice is bandlimited at $L=64$, giving our ground truth mass map, to which we apply a basic galactic plane and ecliptic mask.

Synthetic shear data is generated by applying the measurement operator $\bm{\Phi}$ to the ground truth map.
The noise in weak lensing surveys depends on the number of galaxy counts per unit area and the variance of the intrinsic ellipticity distribution $\sigma_e \sim 0.37$.
We choose an overall number density of galaxies per arcmin$^2$, $n_{\mathrm{gal}}$, and add zero-mean Gaussian noise to the synthetic data, with the variance of the noise given by
\begin{equation}
    \sigma^2_t = \frac{\sigma_e^2}{\sqrt{2n_t}},
\end{equation}
where $t$ is the colatitude index, and $n_t$ is the expected number of galaxies in a pixel at colatitude $t$ given the overall number density $n_{\mathrm{gal}}$.
This dependence on colatitude comes from the equiangular nature of the MW sampling theorem \citep{McEwen2011}.

When reporting the SNR, R2E summary statistics and our solution maps, we use a slightly larger mask than what was applied to the synthetic shear data.
This is to remove leakage artefacts that occur around the edge of the mask due to the sudden lack of data and also wavelets that have support both inside and outside the mask.
At the high resolutions typically of interest for mass-mapping, this leakage will be minimal.

\begin{figure*}
    \includegraphics[width=\textwidth]{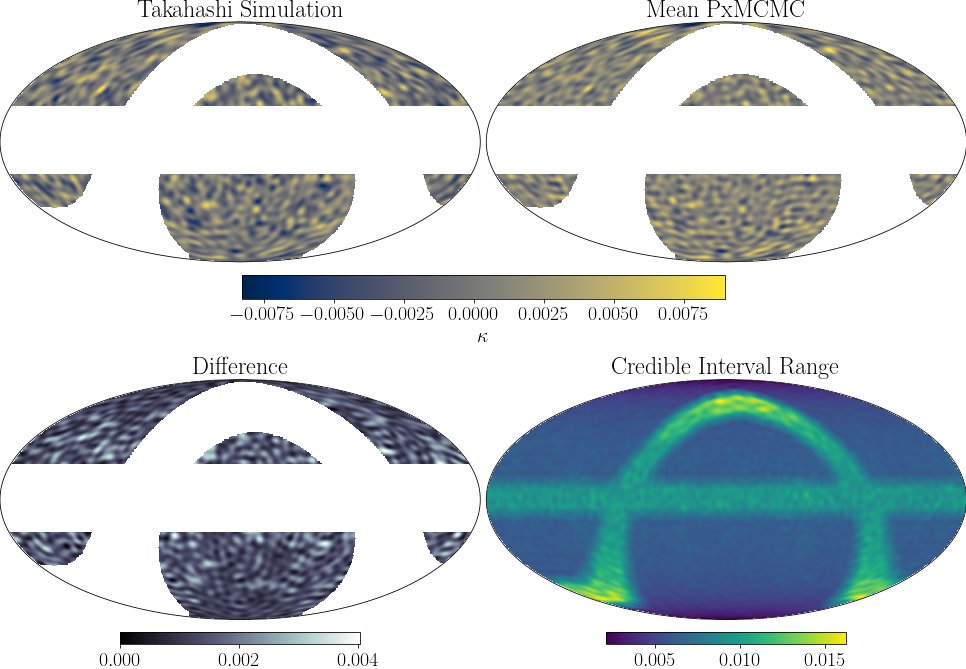}
    \caption{Results of a simple mass-mapping example where the ground truth is known.
    (top left) Ground truth simulation from \citet{Takahashi2017}.
    (top right) Mean of the proximal MCMC samples.
    (bottom left) Difference between the top two maps.
    (bottom right) 95\% credible interval range obtained from proximal MCMC.
    A galactic plane and ecliptic mask is show in all maps except the uncertainty map, which, as expected, shows high uncertainty in these regions.}
    \label{fig:wl_results}
\end{figure*}

Maps of the ground truth, mean of our MCMC samples, the difference between them and our measured uncertainty are shown in Figure~\ref{fig:wl_results}.
For this inversion, we perform $12.5\times10^6$ chain steps, saving every 500\textsuperscript{th} sample and using the last 3000 samples for our results, taking about 4 days on a 2.5 GHz Intel Xeon Platinum 8180M processor.
The step-size parameter $\delta$ is chosen to be $10^{-9}$, the largest value that allows a stable Euler approximation of the Langevin diffusion.
The regularisation parameter $\mu$ is chosen to be $5\times10^5$, to constrain the wavelet coefficients to the appropriate order of magnitude.
At this resolution ($L=64$), there are \num{18916} wavelet coefficients to be sampled.
The mean solution has an SNR of 7.83 dB and a R2E of 0.1.
The uncertainty map, measured as the range of the 95\% credible intervals (equation~\ref{eqn:ci_range}) show, as expected, high uncertainty within the masked regions where the solution is not constrained by any data.

While we obtain encouraging results here with a good reconstruction and physically reasonable pixel-level uncertainties, for this method to be adopted in full-sky mass mapping, computational advances are needed in the implementation of the spherical harmonic and wavelet transforms such that the scaling with $L$ is not as severe.
Currently, the complexity scaling of the spherical harmonic transforms is $\mathcal{O}(L^3)$ \citep{McEwen2011}, and dominates over the efficient harmonic space wavelet transform \citep{Leistedt2013} discussed in Section~\ref{sec:background}.
With at least two spherical harmonic transforms per iteration of MYULA, current implementations will not allow for the desired $L\sim\mathcal{O}(10^3)$ bandlimits.
New efforts implementing these transforms on GPUs should go a significant way towards pushing our method to higher resolutions.
Further theoretical advances for the representation of spherical signals could also lead to computational savings \citep[e.g.][]{Ocampo2022}.

\section{Conclusions}
\label{sec:concl}
In this paper we have presented a general framework for posterior sampling of inverse problems on the sphere with sparsity promoting priors that allows for flexible uncertainty quantification and extends naturally to non-linear problems.
We demonstrated the potential applicability of this method to both astrophysical and geophysical problems.
As with all MCMC methods, the suitability of this method depends on the time taken to take the next chain sample, particularly in relation to the forward modelling step.
The computational complexity of transforms on the sphere mean that this framework is generally feasible for problems of low to moderate resolution (roughly $L\leq64$), such as those commonly considered in global seismic tomography.
At higher resolutions, as needed for full-sky mass-mapping, posterior sampling quickly becomes intractable largely due to the poor scaling of spherical harmonic transforms present in the forward operator.
In either case, special consideration must be given to the forward operator and whether it should be formulated in harmonic or pixel space, and also if its adjoint is known.
A harmonic formulation would avoid repeated spherical harmonic transforms, but these savings could be lost on the measurement operator.
Making use of a more efficient proximal algorithm based on a stochastic Runge-Kutta approximation of the Langevin equation \citep{Pereyra2020} could be a promising way forward for higher resolution spherical inverse problems.
This algorithm is more complex but converges to a solution much faster than the Euler approximation algorithm used in this work, thereby potentially requiring fewer spherical harmonic transforms.
Additionally, faster implementations of the spherical harmonic transforms leveraging GPUs would immediately increase the potential of our method.

The examples shown in this work are largely illustrative, as simplifications have been made.
In the mass-mapping example, the bandlimit is much lower than the angular orders at which the convergence spectrum has the most power due to the computational restrictions imposed by the spherical harmonic transforms.
We also have not fully considered the effect of reduced shear.
Nonetheless, ongoing work to implement the spherical harmonic and wavelet transforms on accelerators (GPUs) provide a route to scale to higher resolutions, making our framework is a promising addition to other recent methods for obtaining mass-maps with uncertainties \citep[e.g.][]{Price2020,Fiedorowicz2022}.
For the surface wave tomography example, we have used the great circle approximation and not accounted for anisotropic effects, in which case a least-squares approach is fast and efficient.
Nonetheless, our results demonstrate the feasibility of our framework methods on global scale tomographic inverse problems.
In this case, the more commonly used harmonic formulation of the forward problem proved to be too slow, and we were able to reformulate it in pixel space such that it was much faster even with the computational overhead of spherical wavelet transforms.
Further, the uncertainties calculated from our posterior samples make physical sense, being correlated with the distribution of data.
Bayesian methods in seismic tomography on large-to-global scales have largely been used for independent 1D inversions \citep[e.g.][]{Shapiro2002,Khan2011,Ravenna2017}, although new advances in gradient-based or variational methods \citep[e.g.][]{Fichtner2018,Gebraad2020,Zhang2020,Zhao2020} are promising for 2D and 3D probabilistic tomography.
Our framework is a further contribution to this advance in methodology, with the novelty of being able to use a non-differentiable prior.

\section*{Acknowledgements}
A.M. is supported by the STFC UCL Centre for Doctoral Training in Data Intensive Science (grant number ST/P006736/1).
A.M.G.F. is grateful to funding from the European Research Council (ERC) to the UPFLOW consolidator grant under the European Union's Horizon 2020 research and innovation programme (grant agreement No 101001601).  
\section*{Data Availability}\label{app:software}
We provide a new Python package, \href{https://github.com/auggiemarignier/pxmcmc}{\texttt{pxmcmc}}\footnote{\url{https://github.com/auggiemarignier/pxmcmc}} that contains implementations of the proximal MCMC algorithms used in this work as well as the measurement operators, wavelet transforms and priors with their proximal mappings.
The code is designed to be flexible, with base classes that will allow users to implement their own forward (measurement and transform) operators and priors.
The proximal MCMC algorithms implemented are themselves not restricted to spherical problems, as the spherical aspects of the inversions appear in the likelihood and prior proximal calculations.
We also provide scripts and data to reproduce the synthetic experiment described in this paper.
MCMC chains for the real data inversions are available from the authors upon request.
The discretisation of great circle paths is implemented in our Python code \href{https://github.com/auggiemarignier/GreatCirclePaths}{\texttt{greatcirclepaths}}\footnote{\url{https://github.com/auggiemarignier/GreatCirclePaths}}, which is publicly available.



\bibliographystyle{rasti}
\bibliography{IEEEabrv,jabbrev,references} 





\bsp	
\label{lastpage}
\end{document}